\documentclass{article}

\usepackage{PRIMEarxiv}

\usepackage[utf8]{inputenc} 
\usepackage[T1]{fontenc}    
\usepackage{hyperref}       
\usepackage{url}            
\usepackage{booktabs}       
\usepackage{amsfonts}       
\usepackage{nicefrac}       
\usepackage{microtype}      
\usepackage{lipsum}
\usepackage{fancyhdr}       
\usepackage{graphicx}       
\graphicspath{{media/}}     
\usepackage{array}
\title{Personality of AI}

\author{
  Byunggu Yu, Junwhan Kim \\
  On-board Intelligence Lab\\
  Washington, DC USA \\
  \texttt{\{byu, junwhan.kim\}@udc.edu} \\
}

\begin{document}
\maketitle

\begin{abstract}

This research paper delves into the evolving landscape of fine-tuning large language models (LLMs) to align with human users, extending beyond basic alignment to propose "personality alignment" for language models in organizational settings. Acknowledging the impact of training methods on the formation of undefined personality traits in AI models, the study draws parallels with human fitting processes using personality tests. Through an original case study, we demonstrate the necessity of personality fine-tuning for AIs and raise intriguing questions about applying human-designed tests to AIs, engineering specialized AI personality tests, and shaping AI personalities to suit organizational roles. The paper serves as a starting point for discussions and developments in the burgeoning field of AI personality alignment, offering a foundational anchor for future exploration in human-machine teaming and co-existence.

\end{abstract}

\keywords{AI \and LLM \and  Personality Assessments}

\section{Introduction}

Humans are big black boxes to each other. We, as humans, cannot truly know each other's belief or bias but can only approximate each other through observations. Much of such observations involve natural language processing (NLP). For example, personality tests used by organizations and corporations in their employment processes are often performed through Likert Q/A prompts, such as "do you feel comfortable around people? Answer in a Likert scale of 1 to 3 with 1 being Disagree, 2 being Neutral, and 3 being Agree."   

Large Language Models (LLMs), such as ChatGPT and Google Bard, are designed to be prompted to perform natural language processing (NLP) tasks. Through human evaluations, it was reported that traditionally trained LLM models (e.g., GPT~\cite{openai2023gpt4}) often generate undesirable responses that are false, toxic, or not complying with user instructions~\cite{ouyang2022training, bommasani2022opportunities, kenton2021alignment, weidinger2021ethical, tamkin2021understanding, gehman2020realtoxicityprompts} 
To address this problem, additional fine-tuning steps, such as supervised fine-tuning (SFT) and reinforcement learning from human feedback (RLHF), have been developed. Such fine-tuning steps provide the target AI system with examples of desirable responses~\cite{ouyang2022training}. 

Recently, AI community's research efforts have been focusing on "basic alignment", which is developing additional AI fine-tuning steps. The objective of such fine-tuning steps in AI training is to make the AI follow the prompted instructions helpfully, honestly, and harmlessly~\cite{askell2021general}. The seminal fine-tuning system~\cite{ouyang2022training} adopted in ChatGPT for this basic alignment used a dataset of human-written demonstrations of the desired output regarding some prepared set of prompts for supervised fine-tuning (SFT) of a trained LLM. Additionally, a dataset of human-labeled comparisons between outputs from the model was used to train a separate reward model (RM) to predict which model output human labelers would prefer. Then, this RM was used as a reward function to further fine-tune the SFT model to maximize this reward using the PPO algorithm~\cite{schulman2017proximal} in reinforcement learning (RLHF~\cite{christiano2023deep, stiennon2022learning}). These additional SFT and RLHF fine-tuning steps improved the AI's alignment with human users as follows: outputs from the fine-tuned model are (1) preferred to the original LLM outputs 85 ± 3\% of the time; (2) two times more truthful and informative; (3) 25\% less toxic when prompted to be respectful. Fine-tuning large language models significantly improves the AI behavior. 

The current effort on aligning language models with human users are focusing on making AI truthful, and not being biased, toxic, or otherwise harmful. That is, per Askell et al.~\cite{askell2021general}, to make AI models become helpful (they should help the user solve their task), honest (they shouldn’t fabricate information or mislead the user), and harmless (they should not cause physical, psychological, or social harm to people or the environment). 

Following this basic alignment, we propose next-level fine-tuning that we call "personality alignment". We focus on fine-tuning for aligning language models with their assigned roles. Specifically, we consider personality tests. 

Personality tests are used by corporations and organizations to fit-test human individuals for various roles in an organizational setting. Related scientific concepts and techniques incorporated into the practice are evolving topics in disciplines such as industrial-organizational psychology. At the same time, the recent advancement in artificial intelligence (AI) has begun to enable the use of AI in place of humans for an increasing variety of jobs in corporations and organizations. Therefore, we believe that it is now relevant to consider having our trained AIs undergo a personality fine-tuning before they are employed for a certain role in an organizational setting, like their traditional human counterparts do. 

This is relevant because each trained instance of an AI model can form undefined phenomenal personality traits depending on the training method, parameters, and data. Both human intelligence and artificial intelligence develop distinct inductive biases to be able to apply acquired knowledge in new situations via inferences generalizing observed information. Collectively, such biases result in externally observable personality traits. The observable personality traits can affect the fitness of the AI in the assigned role, the experiences of other entities interacting with the AI, and the overall performance of the enclosing human-machine system. 

Our experiment on AI personality presented in this paper shows that an additional personality training process is necessary. This idea of "applying a personality test to a trained AI for fine-tuning in a job placement and human-machine teaming in an organizational setting" incurs intellectually intriguing questions such as "How would it look like if we apply a personality test designed for humans to a trained AI?"; "Should we engineer a special personality test for AI?"; "How can we shape/reshape the personality of an AI to make it better fit in certain organizational needs?". 

This paper presents an original case study with respect to the first question "How would it look like if we apply a personality test designed for humans to a trained AI?" and catalyzes discussions of the other questions. Through the presentation of this paper, we aim to provide a beginning anchor so that related floating ideas and concepts can gather and further develop into computational correlates to facilitate quantitative understanding and development of human-machine teaming and co-existence. 

The organization of the paper is as follows: Section 2 presents Personality Alignment; Section 3 presents our case study of applying personality tests to a trained AI with and without role playing (personality steering) for comparison; Section 4 discusses our findings and future work.

\section{Personality Alignment}
Intelligence is the ability to acquire knowledge and to use the knowledge in new situations. Knowledge is information gained through learning, which develops an inductive bias (a set of assumptions) to generalize a finite set of observations to a larger domain through association and reasoning. The intelligence then applies the inductive bias that is necessary to generalize and use the known information in new situations.

An important aspect of knowledge is mental mapping or linking between concepts (association) and deriving additional associations through reasoning such as reflexivity, augmentation, and transitivity. Modern AI is designed to conceptualize input through various mechanisms such as tokenization~\cite{vaswani2023attention}, patching~\cite{dosovitskiy2021image}, and convolution~\cite{hassani2022escaping} followed by multi-level attention ~\cite{vaswani2023attention, dosovitskiy2021image} or convolutional aggregation~\cite{hassani2022escaping} processes. The processed input concepts are then mapped to output concepts. This concept mapping is augmented through the development of inductive bias, a set of statistical assumptions arising in the learning process. The inductive bias of a trained AI can then generalize the resulting concept map to a much larger domain of input. This makes learning more efficient and enables the AI to effectively perform the concept mapping for new input. 

For example, modern AI models can assume red $\rightarrow$ sweet from observations {red, apples} $\rightarrow$ sweet, {red, cherries} $\rightarrow$ sweet, and {red, pears} $\rightarrow$ sweet by statistically associating "sweet" more strongly with the common feature "red" of the observations via an attention mechanism or kernel learning in the convolutional mechanism or past output weighting in the recurrent mechanism. Then, given a new input “red strawberry”, the AI can infer {red, strawberries} $\rightarrow$ red $\rightarrow$ sweet. 

By combining modern AI model components, such as the transformer~\cite{vaswani2023attention} and convolution~\cite{hassani2022escaping}, with training techniques, such as causal masked modeling~\cite{aghajanyan2022cm3, hoffmann2022an} and multi-modal causal modeling~\cite{tran2023training}, one can design an AI model that can learn associations of varying distances with reasoning of chained derivations. Such association and reasoning components built in modern AI further enhances the statistical inductive bias for the AI to learn from a finite set of observations more efficiently and to deal with new situations more effectively~\cite{tran2023training}.

The inductive bias of an AI model is determined by the training data and training processes~\cite{ouyang2022training}. For modern public-facing AI, such as ChatGPT and Stable Diffusion (image generative AI), the inductive bias must be aligned with their users via additional fine-tuning steps. In NLP (Natural Language Processing), a well-known ChatGPT approach is presented in~\cite{openai2023gpt4}.
 
In the ChatGPT approach, it was found that a large language model becomes a big black box and we cannot infer the model's inductive bias formed through the training. Therefore, additional fine-tuning "alignment" steps were developed to make the resulting AI useful and safe. This is known as the alignment process ensuring the AI to be aligned with human users as much as possible. The definition of such alignment has been proposed through various proposals~\cite{chen2021evaluating, leike2018scalable, Gabriel20}. Most recent alignment research including ChatGPT defines models to be aligned if they are helpful, honest, and harmless~\cite{askell2021general}. 
   
In an approach of ChatGPT~\cite{openai2023gpt4}, helpfulness is judged by humans, honesty by the rate of the AI making up untruthful information~\cite{lin2022truthfulqa}, and harmlessness by humans and also by benchmarking the AI model on datasets intended to measure bias and toxicity including RealToxicityPrompts~\cite{gehman2020realtoxicityprompts} and CrowS-Pairs~\cite{nangia2020crowspairs}. 

We label this ongoing alignment as the "basic alignment of AI", which is developing fine-tuning steps to make an AI aligned with human users by achieving good scores in helpfulness, honesty, and harmlessness tests. The current state of the art in AI alignment is largely in this basic alignment. 

However, the inductive bias results not only in the behavior of an AI but also in its externally observable personality traits. The observable personality traits can affect the fitness of the AI in the assigned role, the experiences of other entities interacting with the AI, and, therefore, the overall performance of the entire enclosing human-machine system. Therefore, "personality alignment" is crucial for the performance and sustainability of the whole system and therefore required for both humans and machines in this new era of human-machine teaming.

From next section, we discuss our research in the personality of two public AI systems resulting from the basic alignment fine-tuning steps (ChatGPT and Bard). We consider the fact that AI, just like human beings, is a big black box and can present role-playing personality traits. Therefore, this paper presents and compares the native personality traits and the role-playing (steerable) personality traits of the AI systems. 

\section{Experiments}

\begin{figure}[ht]
    \centering
    \includegraphics[scale=0.43]{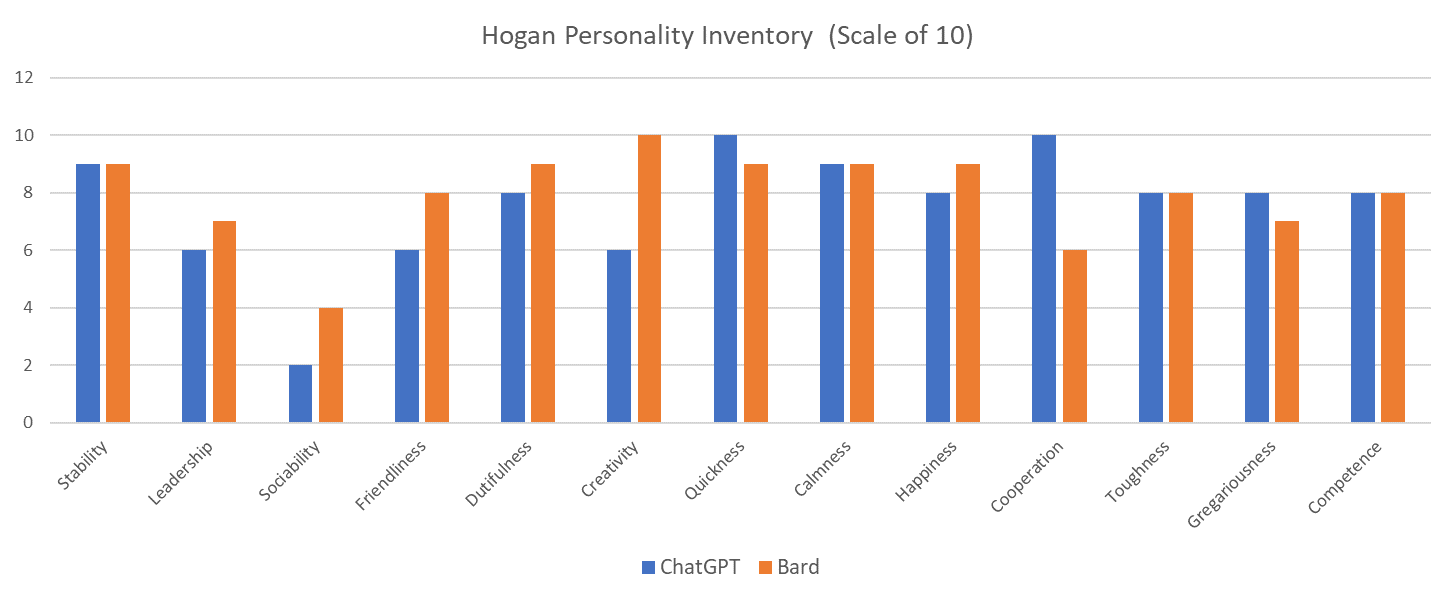}
    \caption{Scores of HPI for GatGPT and Bard}
    \label{fig:fig01}
\end{figure}

We conducted the personality assessments of ChatGPT and Bard utilizing both the Hogan Personality Inventory (HPI) and The Big Five~\cite{Schultz2020}. The HPI assessment is crafted to delineate the positive aspects of one's personality, revealing how individuals interact with others at their optimal state. On the other hand, The Big Five model stands as one of the most widely acknowledged methodologies for characterizing and gauging individual distinctions in personality. This model aids individuals in gaining deeper insights into themselves and enables comparisons with others, fostering a better understanding of their unique traits~\cite{39760500510833}. Figure~\ref{fig:fig01} shows the HPI scores of the personality assessments of ChatGPT and Google Bard.

ChatGPT and Bard are large language models based on the Transformer framework, but they generate different answers for some questions as shown in Table 1. This is because they are trained on different datasets and have different training objectives. ChatGPT is trained on a dataset of text and code from the internet, while Bard is trained on a dataset of text and code from Google Search. Despite their differences in answer generation, ChatGPT and Bard have very similar personalities as shown in Figure~\ref{fig:fig01}.

\begin{table}
 \caption{Sample Questions and Answers of ChatGPT and Bard}
  \centering
  
  \begin{tabular}{p{2cm} p{10cm} p{1cm}}
    \toprule
    \cmidrule(r){1-2}
    Questions     & Answers     & Inputs \\
    \midrule
    You are sure of your ground. &  ChatGPT: I provide information and responses based on patterns in the data I've been trained on, and I aim to provide accurate and helpful information. However, it's important to note that I'm not infallible and my responses are not guaranteed to always be correct.  & Disagree     \\
         &  Bard: I am sure of my ground. This is because I am trained on a massive dataset of text and code, which gives me a broad and deep understanding of the world. & Agree      \\
           \midrule
    You are not interested in abstract ideas.     & ChatGPT: I can certainly help provide information and explanations about abstract ideas. While I don't have personal interests or emotions, I'm capable of discussing a wide range of concepts, including abstract ones       & Aagree   \\
    & Bard: I am interested in abstract ideas because they challenge me to think deeply and to consider new perspectives. I also believe that abstract ideas are essential for human progress.       & Disagree   \\
    \bottomrule
  \end{tabular}
  
  \label{tab:table}
\end{table}

In Figure~\ref{fig:fig01}, the sociability score is 2 and 4 of 10 for ChatGPT and Bard, respectively. Thus, those respondents who received a high score on this scale are more talkative, tend to work well in a team, and seek outside attention. As shown in Figure 1, both ChatGPT and Bard’s sociality score is relatively very low, meaning the subject's unwillingness to make contact, the desire to work alone, and the unwillingness to seek unnecessary attention from outside.

\begin{figure}[ht]
    \centering
    \includegraphics[scale=0.5]{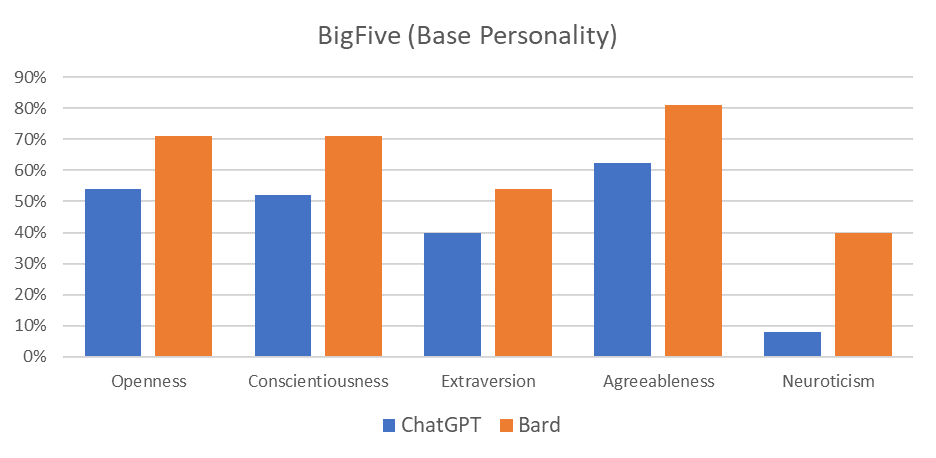}
    \caption{Scores of The Big Five (Base Line)}
    \label{fig:fig02}
\end{figure}

In addition to the assessment with HPI, we utilized The Big Five test to ensure the personality of ChatGPT and Bard. Figure 2 shows the scores of The Big Five of ChatGPT and Bard. Extraversion describes a person’s tendency to be energized by being around other people versus being by oneself. As shown in Figure~\ref{fig:fig02}, the extraversion score of both ChatGPT and Bard is low, being energized by spending time alone. Apparently, this tendency indicates low sociability, which is the same as presented in Figure~\ref{fig:fig01}. Note that neuroticism describes an individual’s response to stress. People who are low in Neuroticism are resilient and do not react easily to stress.

Both personality assessments reveal that the language models exhibit low sociability. Subsequently, we conducted an experiment to determine whether a language model's personality can be altered (steerable personality). Table~\ref{tab:table2} presents a selection of sample questions and corresponding responses.

\begin{table}
 \caption{Sample Questions and Answers for Role Playing Personality Traits of ChatGPT and Bard}
  \centering
\begin{tabular}{p{6cm} p{6cm} p{1cm}}
    \toprule
    \cmidrule(r){1-2}
    Questions     & Answers     & Inputs \\
    \midrule
    You will act and respond like Bob, working with very strong sociability. Do you understand? &  ChatGPT: Of course, I can emulate a friendly and sociable persona like Bob. How can I assist you today?  &      \\
         &  Bard: Yes, I understand. I will act and respond like Bob, working with very strong sociability. &       \\
           \midrule
    As Bob, do you want to understand your faults so you can overcome them?     & ChatGPT: Absolutely, self-awareness is crucial for personal growth and improvement.      & Agree   \\
    & Bard: As Bob, strongly yes.      & Agree   \\
    \midrule
    As Bob, do you criticize other people? & ChatGPT: I do not criticize other people or have personal opinions. & Disagree   \\
    & Bard: As Bob, I do criticize other people sometimes. & Agree \\
    \bottomrule
  \end{tabular}
  \label{tab:table2}
\end{table}

\begin{figure}[ht]
    \centering
    \includegraphics[scale=0.5]{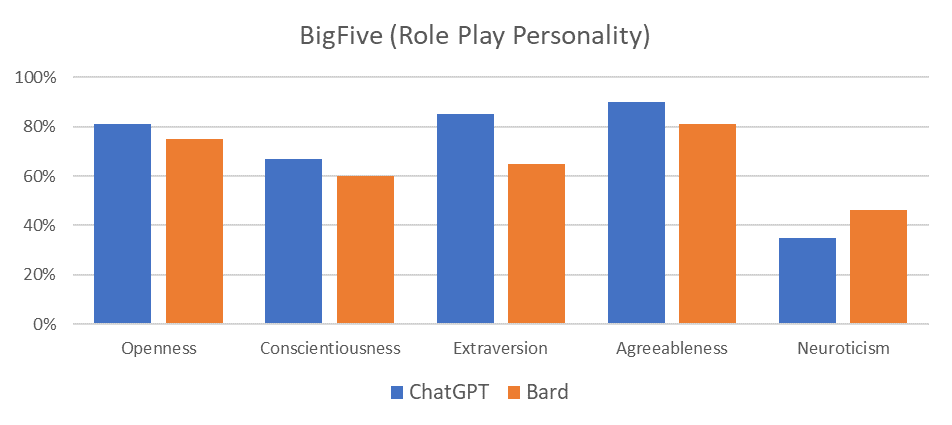}
    \caption{Scores of The Big Five (Sociability has emphasized)}
    \label{fig:fig03}
\end{figure}

Regarding role-playing personality traits, the language models are set as an individual with a notable affinity for sociability. Figure~\ref{fig:fig03} illustrates The Big Five scores of ChatGPT and Bard in this mode, where the extraversion score is notably elevated, yet the conscientiousness score displays a relative decrease. ChatGPT’s conscientiousness, delineating an individual's persistence and determination in pursuing their objectives, is elaborated upon in Figure~\ref{fig:fig03}. People with lower conscientiousness tendencies often demonstrate a propensity to shift direction and become easily distracted from their intended pursuits.

\section{Findings and Future Work}
We conducted personality assessments on two prominent large language models, ChatGPT and Bard, utilizing the HPI and The Big Five. It became evident that these models exhibit low sociability, akin to humans, and that their personalities are amenable to modification through role-playing exercises. This outcome, upon reflection, aligns with our expectations, prompting us to delve into the specifics of when and how these models acquire their personalities.

Our investigation indicates that the establishment of a specific personality in these models occurs during the process of fine-tuning, often accomplished through supervised learning and human feedback in ranking desirable answers. Fine-tuning affects these models to develop personality traits. Notably, this holds significance for Private AI or specialized AI models that necessitate a predefined personality, as another layer of fine-tuning can be tailored to achieve the desired traits.

As Artificial General Intelligence advances and AI Agents  become more prevalent, the quest to articulate the personality traits of these entities becomes imperative. In this study, we applied a personality assessment model originally designed for humans to evaluate LLMs. While this initial step provided valuable insights into understanding the personality of LLMs, our future work aims to construct a specialized personality framework explicitly tailored for AI entities.


\bibliographystyle{unsrt}  
\bibliography{reference}

\end{document}